\documentclass[11pt,twoside]{jgsp}

\newcommand{\ri}{\mathrm{i}}
\newcommand{\rd}{\mathrm{d}}

\usepackage{bm}

\def\Tr{
{\rm Tr}}

\def\det{
{\rm det}}

\def\const{
{\rm const}}

\theoremstyle{plain}

\def\aut#1#2{
{

\noindent
\parbox[t]{6.5cm}{#1}
 \hfill
\parbox[t]{6.5cm}{#2}
}
}

\DeclareMathOperator{\Complex}{\mathbb{C}}
\DeclareMathOperator{\Integer}{\mathbb{Z}}
\DeclareMathOperator{\Jac}{Jac}
\DeclareMathOperator{\Symm}{Symm}
\DeclareMathOperator{\Ibb}{\mathbb{I}}
\DeclareMathOperator{\pr}{pr}

\begin{document}

\thispagestyle{plain}

\title{Harmonic analysis on Lagrangian manifolds \\ of integrable Hamiltonian systems}
\author{Julia Bernatska and Petro Holod}

\date{}

\maketitle

\comm{Communicated by Alexandar B. Yanovski}\

\begin{abstract}
For an integrable Hamiltonian system we construct a representation
of the phase space symmetry algebra over the space of functions on a Lagrangian manifold.
The representation is a result of the canonical quantization of the integrable system
in terms of separation variables. The variables are chosen in such way that a half of them
parameterizes the Lagrangian manifold, which coincides with the Liouville torus of the integrable system.
The obtained representation is indecomposable and non-exponentiated.
\end{abstract}

\label{first}
\section[]{Introduction}
The problem of quantization on a Lagrangian manifold has arisen from the theory
of geometric quantization \cite{Karasev}. But the question how to choose a
proper Lagrangian manifold remains open. Dealing with a dynamical system we use its
Liuoville torus as a Lagrangian manifold. This choice guarantees that
the representation space consists of holomorphic functions - functions on
the special Lagrangian manifold whose complexification serves as a phase space of the system.

According to the orbit method one can construct an intergable soliton hierarchy (hierarchy of equations of soliton type) on orbits of a loop group \cite{Holod}. Finite gap phase spaces for the intergable hierarchy is appeared to consist of orbits of finite quotient algebras corresponding to the loop group.
On such phase space one can introduce canonical variables of separation (Darbu coordinates), which represent points of a spectral curve \cite{BernatskaJNMP}. The curve is hyperelliptic for many interesting intergable systems. A half of the variables of separation parametrizes the Lagrangian manifold which is the Liouville torus for the intergable system in question, and the complexified Lagrangian manifold serves as a generalized Jacobian of the spectral curve.

Canonical quantization in terms of the variables of separation gives rise to a representation for the symmetry group of the phase space. We construct such representation in the space of holomorphic functions on the complexified Lagrangian manifold, and perform a harmonic analysis of the representation
for the system of isotropic Landau-Lifshits equation (for a finite gap phase space).

\section[]{Preliminaries}
We deal with systems on orbits of the loop algebra
$\widetilde{\mathfrak{g}}\,{=}\,\mathfrak{sl}(2,\Complex)\,{\times}\,\mathcal{P}(z,z^{-1})$.
In particular, on these orbits one can construct the integrable heirarchies of modified Korteweg-de Vries equation,
sin(sh)-Gordon equation, nonlinear Schr\"{o}dinger equation, and iso\-tro\-pic Landau-Lifshits equation,
for more details see \cite{BernatskaJNMP}. The systems obey the Lax equation
\begin{gather*}
\frac{\rd L(z)}{\rd t} = [A(z),L(z)],\qquad
\widetilde{\mathfrak{g}}^{\ast} \ni
L(z) = \begin{pmatrix} \alpha(z) & \beta(z) \\
\gamma(z) & -\alpha(z) \end{pmatrix}\\
\alpha(z) = \sum_{j=0}^{N} \alpha_j z^j,\quad
\beta(z) = \sum_{j=0}^{N} \beta_j z^j, \quad
\gamma(z) = \sum_{j=0}^{N} \gamma_j z^j,
\end{gather*}
where $\alpha_N$, $\beta_N$, $\gamma_N$ are constant.
The matrix $A \,{\in}\, \widetilde{\mathfrak{g}}$ defines a heirarchy.
For example, the hierarchy of Landau-Lifshits equation is
obtained by  means of
$$A(z) = -\frac{1}{z}
\begin{pmatrix} \alpha_1 & \beta_1 \\
\gamma_1 & -\alpha_1 \end{pmatrix}
- \frac{1}{z^2}
\begin{pmatrix} \alpha_0 & \beta_0 \\
\gamma_0 & -\alpha_0 \end{pmatrix}. $$

\subsection[]{Phase Space of the Integrable System}
According to the Kostant-Adler scheme, the coadjoint action of finite quotient algebra
$\mathfrak{g}\,{\times}\,\mathcal{P}(z^\nu,\dots,z^{\nu + N-1})$ over
the finite subspace $\mathcal{M}^{N}\,{\equiv}\,\mathfrak{g}^{\ast}\,{\times}\,
\mathcal{P}(1,\,z,\,\dots$, $z^{N})$
of $\widetilde{\mathfrak{g}}^{\ast}$  produces
a set of orbits $\mathcal{O}^{N}\,{\in}\,\mathcal{M}^{N}$, which serves as
an $N$-gap phase space of an integrable system.
Choosing different $\nu$, one can construct different Hamiltonian systems
generated by a series of Poisson structures.

The Lax equation guarantees that evolution of
a system preserves the spectrum of matrix $L$.
Thus the quantities $\Tr L^k$ are automatically constants of motion, and one gets as many
as the order of $L$. A half of these constants defines an orbit $\mathcal{O}^{N}$,
the rest forms a complete set of integrals of motion, which we call Hamiltonians.

All such systems are algebraic integrable, that is
integrable in Kowalewska sense: every solution of the system admits a holomorphic continuation in time.
So every solution is associated with a Riemannian surface
$\mathcal{R}$. The constant spectrum provides existence of a spectral curve,
which is usually defined by the equation
$$\det (L(z)-w)=0.$$
The spectral curve serves as a Riemannian surface $\mathcal{R}$
from the definition of integrability in Kowalewska sense.

As mentioned above, the orbits $\mathcal{O}^N$ form the phase space of an integrable system.
On the other hand, the phase space is the Abelian torus
arising as a complexification of the Liouville torus of the system.
The complexified Liouville torus coincides with a generalized Jacobian
of the mentioned Reimannian surface $\mathcal{R}$:
$$\widetilde{\Jac}(\mathcal{R}) = \Symm\limits_N
\underbrace{\mathcal{R}\times\mathcal{R}\times\cdots\times\mathcal{R}}_{N},\qquad N>g,$$
where $g$ is the genus of $\mathcal{R}$.
The necessity of generalization arises in hierarchies of soliton type equations because
the number $N$ of gaps is usually greater than $g$, see \cite{Previato} for finite gap systems
of the nonlinear Schr\"{o}dinger hierarchy.

\subsection[]{Separation of Variables and Quantization}
Original variables in the phase space are coefficients of the polynomials
$\gamma$, $\beta$, $\alpha$ which are the entries of matrix $L$.
The set of coefficients $\{\gamma_j\,{;}\,j\,{=}\,0,\dots,\,N\}$
are eliminated by means of orbit equations.
So $\{\beta_j,\, \alpha_j\,{;}\,j\,{=}\,0,\dots,\,N\,{-}\,1\}$ serve as independent variables, and
normally they are not canonically conjugate.

In order to construct a Lagrangian manifold, it is suitable to
find conjugate variables. We use the scheme from \cite{BernatskaJNMP}, its idea is the following.
Let $\{z_k,\,w_k\,{;}$ $k\,{=}\,1$, $\dots$, $N\}$  be a set of variables of separation.
If one requires every conjugate pair $(z_k,\,w_k)$ be a point of the spectral curve, then
$\{z_k\}$ should be the roots of polynomial $\beta$.

The proposed scheme enables to construct
variables of separation. Then we define a Lagrangian manifold as
the submanifold parameterized by $\{z_k\,{;}\,k\,{=}\,1,\dots,\,N\}$  (all $w_k$ are fixed),
it coincides with the Liouville torus of the system in question.

Quantization in the Schr\"{o}dinger picture
$$z_k \mapsto \hat{z}_k,\quad
w_k\mapsto\hat{w}_k\,{=}\,{-}\ri \frac{\partial }{\partial z_k},\quad
\{z_k,w_l\}\,{=}\,\delta_{kl} \mapsto [\hat{z}_k,\hat{w}_l]\,{=}\,\ri \delta_{kl} \Ibb$$
in a very natural way gives a representation of the algebra corresponding to the phase space symmetry group,
which we call the phase space symmetry algebra. The obtained algebra representation
is realized by differential operators of high order (higher than one),
and so can not be exponentiated to a group.
It happens because we restrict the domain of functions from the phase space
to a Lagrangian manifold.
This is the difference from the standard geometric quantization.

\section{The Integrable System of Isotropic Landau-Lifshits Equation}

Here we consider the 2-gap system from the hierarchy of isotropic Landau-Lifshits equation,
also called the continuous Heisenberg magnetic chain:
\begin{equation}\label{LandauEq}
\frac{\partial \bm{\mu}}{\partial t} = \frac{1}{2c_0} \left[\bm{\mu},
\frac{\partial^2 \bm{\mu}}{\partial x^2}\right] + \frac{c_1}{2c_0} \frac{\partial \bm{\mu}}{\partial x},
\end{equation}
where the vector $\bm{\mu}$ describes a magnetization, $c_0$, $c_1$ are constants.

\subsection{Phase Space, $\mathfrak{e}(3)$ Structure}
The Lax matrix $L$ looks as follows
\begin{gather*}
L(z) = \begin{pmatrix} \mathrm{i}\mu_3(z) & \mu_1(z) - \mathrm{i}\mu_2(z) \\
-\mu_1(z) - \mathrm{i}\mu_2(z) & -\mathrm{i}\mu_3(z) \end{pmatrix}\\
\mu_{1,2}(z) = \sum_{j=0}^{N-1} \mu^{(j)}_{1,2} z^j,\quad
\mu_{3}(z) = \frac{1}{2}\, z^N + \sum_{j=0}^{N-1} \mu^{(j)}_{3} z^j.
\end{gather*}
The vector $\big(\mu_1^{(0)},\,\mu_2^{(0)},\,\mu_3^{(0)}\big)\,{=}\,\bm{\mu}$
obeys the Landau-Lifshits equation \eqref{LandauEq}.
In the case of 2-gap system ($N\,{=}\,2$) one has
\begin{gather*}
\begin{array}{ll}
& \mu_1(z) = \mu_1^{(0)} + \mu_1^{(1)} z \\
& \mu_2(z) = \mu_2^{(0)} + \mu_2^{(1)} z \\
& \mu_3(z) = \mu_3^{(0)} + \mu_3^{(1)} z + z^2/2.
\end{array}
\end{gather*}
The coefficients $\{\mu_{1,2,3}^{(0)},\,\mu_{1,2,3}^{(1)}\}$ serve as \emph{dynamic variables}, they
form a phase space, which we equip with the Poisson structure
\begin{equation}\label{PoissonStr}
\{\mu_k^{(0)},\mu_l^{(0)}\}=0,\quad \{\mu_k^{(0)},\mu_l^{(1)}\}=\varepsilon_{klj} \mu_j^{(0)},\quad
\{\mu_k^{(1)},\mu_l^{(1)}\}=\varepsilon_{klj} \mu_j^{(1)}.
\end{equation}
This is $\mathfrak{e}(3)$ algebra structure, therefore
the Euclidian group $\text{E}(3)$
serves as a \emph{phase space symmetry group} of the system.
We also call $\mathfrak{e}(3)$  the \emph{phase space symmetry algebra}.

Invariance of the matrix $L$ spectrum provides constants of motion: $h_0$, $h_1$, $h_2$, $h_3$
obtained from the equation
\begin{gather*}
\const = -\Tr L^2(z) = z^4/4 + h_3 z^3 + h_2 z^2 + h_1 z + h_0\\
\begin{array}{lll}
h_0 = (\bm{\mu}^{(0)},\bm{\mu}^{(0)})&\quad& \bm{\mu}^{(0)} \equiv (\mu_1^{(0)},\mu_2^{(0)},\mu_3^{(0)})\\
h_1 = 2(\bm{\mu}^{(1)},\bm{\mu}^{(0)})&\quad& \bm{\mu}^{(1)} \equiv (\mu_1^{(1)},\mu_2^{(1)},\mu_3^{(1)})\\
h_2 = (\bm{\mu}^{(1)},\bm{\mu}^{(1)}) + \bm{\mu}^{(0)}_3\\
h_3 = \bm{\mu}^{(1)}_3.
\end{array}
\end{gather*}
The functions $h_0$, $h_1$ annihilate the Lie-Poisson bracket,
they define an orbit $\mathcal{O}$:
$$h_0 = c_0,\qquad h_1 = c_1,$$
where $c_0$ and $c_1$ are arbitrary constants.
The functions $h_2$, $h_3$ serve as integrals of motion
called Hamiltonians.

The spectral curve, which is the Riemannian surface $\mathcal{R}$, is of genus 2:
$$z^4 w^2 = z^4/4 +
h_3 z^3 + h_2 z^2 + c_1 z + c_0.$$

In what follows we change notations from $\bm{\mu}^{(0)}$ and
$\bm{\mu}^{(1)}$ to $\bm{p}$ and $\bm{L}$ vectors:
\begin{gather*}
\bm{\mu}^{(0)} \equiv \bm{p},\qquad  \bm{\mu}^{(1)} \equiv \bm{L}.
\end{gather*}
Then the orbit equations get the form
\begin{subequations}\label{OrbEqs}
\begin{gather}
\bm{p}^2 = c_0  \label{OrbSphere} \\
(\bm{p},\bm{L}) = c_1/2. \label{OrbPlane}
\end{gather}
\end{subequations}
Evidently, the orbit is a bundle of the planes \eqref{OrbPlane}
over the sphere \eqref{OrbSphere}: the plane is attached to every point $\bm{p}$ of the sphere.
Using different values of $c_0$ and $c_1$ one obtains a set of orbits.
All such orbits form the phase space of the system.
There exists a degenerate orbit collapsed into the point $\bm{p}\,{=}\,0$, that corresponds to
the case $c_0\,{=}\,0$, $c_1\,{=}\,0$.

In the new notations the Hamiltonians look as follows
$$h_2= \bm{L}^2 + p_3,\qquad
h_3=  L_3.$$

\subsection{Canonical Quantization}
In order to obtain a representation of the phase space symmetry algebra
we use the canonical quantization (see Preliminaries).
By separation of variables we prepare the system for the quantization, which
gives a representation over the space of functions on the Lagrangian manifold
formed by a half of conjugate variables.

Variables of separation are obtained in the following way, for more details see~\cite{BernatskaJNMP}.
According to the scheme, the variables $z_1$, $z_2$ are roots of the polynomial
$\beta$. But this is a polynomial of degree 1 in our case. The situation is improved by means of the
similarity transformation
$$P^{-1}L(z) P = \begin{pmatrix} \ri\mu_2(z) & \mu_1(z)+\ri\mu_3(z)\\
-\mu_1(z)+\ri\mu_3(z) & -\ri\mu_2(z) \end{pmatrix},\quad
P = \frac{1}{\sqrt{2}}\begin{pmatrix} 1&-1 \\ 1&1\end{pmatrix}.$$
Now the polynomial
$\mu_1(z) \,{+}\, \ri \mu_3(z)$ has two roots: $z_1$,
$z_2$. The conjugate variables are calculated by the formula $w_k \,{=}\, \ri \mu_2(z_k)/z_k^2$.
Explicit expressions for all dynamic variables are given below
\begin{align*}
&p_1 = \ri\left(\frac{z_1z_2}{4} - \frac{c_0}{z_1z_2}-
\frac{z_1z_2(z_1 w_1-z_2 w_2)^2}{(z_1-z_2)^2}\right)\\
&p_2 = \ri z_1 z_2 \frac{z_1 w_1-z_2 w_2}{z_1-z_2}\\
&p_3 = \frac{z_1z_2}{4} + \frac{c_0}{z_1z_2}+
\frac{z_1z_2(z_1 w_1-z_2 w_2)^2}{(z_1-z_2)^2}\\
&L_1 = \ri \left(-\frac{z_1+z_2}{4} -
\frac{c_1}{z_1z_2} - \frac{c_0(z_1+z_2)}{z_1^2z_2^2}+
\frac{z_1^2 w_1^2 - z_2^2 w_2^2}{z_1-z_2}\right)\\
&L_2 = -\ri\frac{z_1^2 w_1-z_2^2 w_2}{z_1-z_2}\\
&L_3 =-\frac{z_1+z_2}{4} +
\frac{c_1}{z_1z_2} + \frac{c_0(z_1+z_2)}{z_1^2z_2^2}-
\frac{z_1^2 w_1^2 - z_2^2 w_2^2}{z_1-z_2}.
\end{align*}

After the canonical quantization:
$z_k \,{\mapsto}\, \hat{z}_k$,
$w_k\,{\mapsto}\,\hat{w}_k\,{=}\,{-}\ri \partial /\partial z_k$
and checking commutation relations we come to a representation
of $\mathfrak{e}(3)$. We write the algebra in the form
\begin{gather*}
\mathfrak{e}(3)=
\{\hat{L}_3,\ \hat{L}_{\pm} = \hat{L}_1\pm \mathrm{i}\hat{L}_2,\ \hat{p}_3,\
\hat{p}_{\pm} = \hat{p}_1\pm \mathrm{i}\hat{p}_2\}\\
[\hat{L}_3,\hat{L}_{\pm}] = \pm \hat{L}_{\pm},\quad [\hat{L}_+,\hat{L}_-] = 2\hat{L}_3,\quad
[\hat{p}_3,\hat{p}_{\pm}] = 0,\quad [\hat{p}_+,\hat{p}_-] = 0\\
[\hat{L}_3,\hat{p}_{\pm}] = [\hat{p}_3,\hat{L}_{\pm}] = \pm \hat{p}_{\pm},\quad [\hat{L}_+,\hat{p}_-] = [\hat{p}_+,\hat{L}_-] = 2\hat{p}_3.
\end{gather*}
The representation of $\mathfrak{e}(3)$ is the following:
\begin{align*}
&\hat{L}_3 = \frac{z_1^2}{z_1-z_2}
\left(\frac{\partial^2}{\partial z_1^2} -\frac{1}{4} -
\frac{c_1}{z_1^3} -
\frac{c_0}{z_1^4}\right) -
\frac{z_2^2}{z_1-z_2}
\left(\frac{\partial^2}{\partial z_2^2} -\frac{1}{4} -
\frac{c_1}{z_2^3} -
\frac{c_0}{z_2^4}\right)\\
&\hat{L}_{\pm} = \frac{\mathrm{i}z_1^2}{z_1-z_2}
\left(-\frac{\partial^2}{\partial z_1^2} -\frac{1}{4} +
\frac{c_1}{z_1^3} +
\frac{c_0}{z_1^4}\mp
\frac{\partial }{\partial z_1}\right) - \\
&\phantom{\hat{L}_{\pm} = \frac{\mathrm{i} z_1^2}{z_1-z_2}
-\frac{\partial^2}{\partial z_1^2} -\frac{1}{4} +
\frac{c_1}{z_1^3}}
\frac{\mathrm{i}z_2^2}{z_1-z_2}
\left(-\frac{\partial^2}{\partial z_2^2} -\frac{1}{4} +
\frac{c_1}{z_2^3} +
\frac{c_0}{z_2^4}\mp
\frac{\partial }{\partial z_2}\right)\\
&\hat{p}_3 = -\frac{z_1 z_2}{(z_1 - z_2)^2}
\left(z_1^2\frac{\partial^2}{\partial z_1^2} + z_2^2
\frac{\partial^2}{\partial z_2^2} - 2z_1z_2
\frac{\partial^2}{\partial z_1 \partial z_2}\right) +\frac{z_1 z_2}{4}+\frac{c_0}{z_1z_2} +
\\ & \phantom{\hat{p}_3 =\frac{z_1 z_2}{(z_1 - z_2)^2}\frac{z_1 z_2}{(z_1 - z_2)^2}
\frac{z_1 z_2}{(z_1 - z_2)^2}\frac{z_1 z_2}{(z_1 - z_2)^2}}
+ \frac{2z_1^2 z_2^2}{(z_1 - z_2)^3}
\left(\frac{\partial}{\partial z_1} - \frac{\partial}{\partial z_2}\right)\\
&\hat{p}_{\pm} = \mathrm{i}\left[\frac{z_1 z_2}{(z_1 - z_2)^2}
\left(z_1^2\frac{\partial^2}{\partial z_1^2} + z_2^2
\frac{\partial^2}{\partial z_2^2} - 2z_1z_2
\frac{\partial^2}{\partial z_1 \partial z_2}\right) +\frac{z_1 z_2}{4}-\frac{c_0}{z_1z_2} - \right.
\end{align*}
\begin{align*}
& \left.\phantom{\hat{p}_3 =\frac{z_1 z_2}{(z_1 - z_2)^2}}
- \frac{2z_1^2 z_2^2}{(z_1 - z_2)^3}
\left(\frac{\partial}{\partial z_1} - \frac{\partial}{\partial z_2}\right)
\pm \frac{z_1 z_2}{z_1-z_2}
\left(z_1 \frac{\partial }{\partial z_1}-z_2 \frac{\partial }{\partial z_2}\right) \right].
\end{align*}
One can easily see that $L_{\pm}$ and $L_3$ admit separation of variables, but
$p_{\pm}$ and $p_3$ have not a good structure for separation.

Then we calculate the Hamiltonians, which also fit separation of variables:
\begin{equation*}
\begin{split}
& \hat{h}_2 = -\frac{z_1^2 z_2}{z_1-z_2}
\left(\frac{\partial^2 }{\partial z_1^2} - \frac{1}{4} - \frac{c_0}{z_1^4} -
\frac{c_1}{z_1^3}\right)+\frac{ z_1 z_2^2}{z_1-z_2}
\left(\frac{\partial^2 }{\partial z_2^2} - \frac{1}{4} - \frac{c_0}{z_2^4} -
\frac{c_1}{z_2^3}\right)\\
&\hat{h}_3 = \frac{z_1^2}{z_1-z_2}
\left(\frac{\partial^2 }{\partial z_1^2} - \frac{1}{4} - \frac{c_0}{z_1^4} -
\frac{c_1}{z_1^3}\right) - \frac{z_2^2}{z_1-z_2}
\left(\frac{\partial^2 }{\partial z_2^2} - \frac{1}{4} - \frac{c_0}{z_2^4} -
\frac{c_1}{z_2^3}\right).
\end{split}
\end{equation*}

The obtained representation of $\mathfrak{e}(3)$ is realized by differential
operators of the second order, therefore it can not be exponentiated to a group.
This is a representation over the space of smooth symmetric functions on
the Lagrangian manifold.

\section{Representation and Harmonic Analysis}
Now we come to a harmonic analysis, which we develop with respect
to the subalgebra $\mathfrak{sl}(2)\,{\subset}\,\mathfrak{e}(3)$.
Firstly we consider the simplest case of degenerate orbit, collapsed into a point:
$$\bm{p}^2=0,\qquad (\bm{p},\bm{L})=0.$$
Its spectral curve $\mathcal{R}$ is reduced to genus 1:
$z^2 w^2 \,{=}\,  z^2/4 \,{+}\,
h_3 z \,{+}\, h_2.$
As a result the operators $\hat{L}_3$, $\hat{L}_{\pm}$ decompose to
the corresponding one-particle operators.
Then we investigate the case of a generic orbit
\begin{gather*}
\bm{p}^2 = c_0 , \qquad
(\bm{p},\bm{L}) =  c_1 /2.
\end{gather*}
We construct a representation space and obtain conditions of quantization.

\subsection{Degenerate Orbit: Representation Space}
We start from an action of
$\mathfrak{sl}(2) \,{=}\, \{\hat{L}_+$, $\hat{L}_-$, $\hat{L}_3\}$.
Thus, we solve the equation
\begin{equation}\label{L3EigenF}
\hat{L}_3 f(z_1,z_2) = m f(z_1,z_2)
\end{equation}
by the method of separation of variables: $f(z_1,z_2) \,{=}\, W_1(z_1)W_2(z_2)$.
Both functions $W_1$, $W_2$ obey the same equation
\begin{gather*}
W'' + \left(-\frac{1}{4} -\frac{m}{z} - \frac{C}{z^2}\right)W = 0,
\end{gather*}
which is the Whittaker equation with $C\,{=}\,\mu^2 \,{-}\,1/4$ and solutions $W_{-m,\mu}$.
If $C\,{=}\,m(m\,{+}\,1)$, the function $f$ also serves as an eigenfunction of $\hat{L}^2$.
We fix a value of $m$ and denote it by $J$, then $\mu\,{=}\,{\pm}(J\,{+}\,1/2)$.
At $\mu\,{=}\,{-}(J\,{+}\,1/2)$ the Whittaker function has a very simple form:
$W_{-J,-J-1/2}(z) \,{=}\, z^{-J} e^{-z/2}.$ This brings to the function
\begin{gather}
f_{JJ}(z_1,z_2) = (z_1 z_2)^{-J} e^{-(z_1+z_2)/2}\label{HighFunc}\\
\hat{L}_3 f_{JJ} = J f_{JJ},\qquad \hat{L}^2 f_{JJ} = J(J+1) f_{JJ}, \nonumber
\end{gather}
which is annihilated by $\hat{L}_+$.
We obtain the highest weight vector
of the $\mathfrak{sl}(2)$ \emph{Verma module} $\mathcal{M}^J$ produced
by the action of $\hat{L}_-$:
\begin{gather*}
f_{Jm}(z_1,z_2) = \ri^{J-m} (J-m)! (z_1 z_2)^{-J} e^{-(z_1+z_2)/2}
\mathcal{L}_{J-m}^{-2J-1}(z_1+z_2)\\ m=J,\,J-1,\,\dots,
\end{gather*}
where $\mathcal{L}_n^{\alpha}$ denotes an associated Laguerre polynomial.
Using the known formula
\begin{equation}\label{ThrAdd}
\mathcal{L}_n^{\alpha}(z_1)\mathcal{L}_n^{\alpha}(z_2) = \sum_{k=0}^n (\alpha+k+1)\cdots(\alpha+n)
\frac{(z_1 z_2)^k}{k!} \mathcal{L}_{n-k}^{\alpha+2k}(z_1+z_2)
\end{equation}
one can expand every function $f_{Jm}$ into a sum of products $W_{-m,\mu}(z_1)W_{-m,\mu}(z_2)$
over $\mu$ from ${-}(J\,{+}\,1/2)$ to ${-}(m\,{+}\,1/2)$, that accords with the variable separation method.

The algebra $\{\hat{L}_+$, $\hat{L}_-$, $\hat{L}_3\}$
acts in the following way:
\begin{equation*}
\hat{L}_3 f_{Jm} = mf_{Jm},\ \ \hat{L}_- f_{Jm} = f_{J,m-1},
\ \ \hat{L}_+ f_{Jm} = (J-m)(J+m+1)f_{J,m+1}.
\end{equation*}
The obtained Verma module has the invariant subspace $\mathcal{M}^{-J-1}$ with the highest weight
vector $f_{J,-J-1}$. Thus, a representation over the quotient
$\mathcal{V}\,{=}\,\mathcal{M}^J\backslash \mathcal{M}^{-J-1}$ is irreducible.

\subsection{Degenerate Orbit: `Unitarization' of $\mathfrak{sl}(2)$ Representation}
The obtained representation is \emph{not canonical}.
Reduction to a canonical representation we call `unitarization',
because normally this procedure brings to a unitary group.
On account of inability to exponentiate the proposed representation
we use quotation marks.

A canonical representation can be constructed by means of the intertwining operator
$\hat{A}$ defined as follows:
\begin{multline*}
\tilde{f}_{Jm} \,{\equiv}\, \hat{A} f_{Jm} =
\sqrt{\frac{\Gamma(J+m+1)}{\Gamma(J-m+1)}}\, f_{Jm} = \\ =
 \ri^{J-m} \sqrt{\Gamma(J+m+1) \Gamma(J-m+1)}
(z_1 z_2)^{-J} e^{-(z_1+z_2)/2}\mathcal{L}_{J-m}^{-2J-1}(z_1+z_2).
\end{multline*}
Indeed, one easily checks that $\mathfrak{sl}(2)$ algebra
has the canonical action:
$$\hat{L}_{\pm} \tilde{f}_{Jm} = \sqrt{(J\mp m)(J\pm m+1)}\, \tilde{f}_{J,m\pm 1},\quad
\hat{L}_3 \tilde{f}_{Jm} = m\tilde{f}_{Jm}.$$
Also we make the basis $\{\tilde{f}_{Jm}\,{;}\, {-}J\,{\leqslant}\,m\,{\leqslant}\, J,\,J\,{=}\,0,\,1,\,\dots\}$
orthonormal by introducing the inner product
\begin{multline*}
\langle \tilde{f}_{Jm}, \tilde{f}_{Jn}\rangle = \int_0^{\infty} \int_0^{\infty}
\frac{\tilde{f}^{\ast}_{Jm}(z_1,z_2) \tilde{f}_{Jn}(z_1,z_2)}
{\Gamma(J-m+1)\Gamma(J+m+1)}
\times \\ \times \frac{\rd z_1 \rd z_2}{z_1^{-J+1}z_2^{-J+1}
\sum_{i=0}^{J-n} \frac{\Gamma(-J+i)}{i!} \frac{\Gamma(-n-i)}{(J-n-i)!}}  = \delta_{nm}.
\end{multline*}
Here we use the summation theorem and the orthogonal relation from \cite{Abramowitz}.
One can observe that `unitarization' by means of the intertwining operator is equivalent to
the Shapovalov formula \cite{Shapovalov}.

\subsection{Degenerate Orbit: Action of $\hat{p}_3$, $\hat{p}_{\pm}$}
With respect to the canonical  representation one gets the following
action of the operators $\hat{p}_3$, $\hat{p}_{\pm}$:
\begin{align*}
&\hat{p}_+ \tilde{f}_{Jm} = - \ri\sqrt{(J-m)(J-m-1)}\, \tilde{f}_{J-1,m+1} \\
&\hat{p}_3 \tilde{f}_{Jm} = -\ri\sqrt{(J-m)(J+m)}\, \tilde{f}_{J-1,m}\\
&\hat{p}_- \tilde{f}_{Jm} = \ri\sqrt{(J+m)(J+m-1)}\, \tilde{f}_{J-1,m-1},
\end{align*}
which matches with the abstract action formulas for $\mathfrak{e}(3)$.

\subsection{Generic Orbit: Representation Space}
In the similar way we deal with a generic orbit.

Again we start with the equation \eqref{L3EigenF}, and come to
a more complicate equation for the functions $W_1$, $W_2$:
\begin{gather}\label{WittEqGen}
W'' + \left(-\frac{1}{4}
- \frac{m}{z} - \frac{C}{z^2} -
\frac{c_1}{z^3} - \frac{c_0}{z^4} \right)W = 0.
\end{gather}
Requiring $\hat{L}_+ W(z_1)W(z_2)\,{=}\,0$, we find the following solution of \eqref{WittEqGen}:
$$W(z) = z^{-m} e^{-z/2+a/z}$$
with an arbitrary $a$. In order to make this function an eigenfunction of $\hat{L}^2$
we should assign $C\,{=}\,J(J\,{+}\,1)\,{+}\,a$, $c_0\,{=}\,a^2$,
$c_1\,{=}\,2a(J\,{+}\,1)$, we again use $J$ for the highest value of $m$.
Then the highest weight vector has the form
$$f_{JJ}(z_1,z_2) = (z_1 z_2)^{-J} e^{-(z_1+z_2)/2+a/z_1+a/z_2}.$$

By the action of $\hat{L}_-$ we produce the $\mathfrak{sl}(2)$ \emph{Verma module}
$\mathcal{M}^J$
\begin{gather*}
f_{Jm}(z_1,z_2) = \ri^{J-m} (J-m)! (z_1 z_2)^{-J} e^{-(z_1+z_2)/2+a/z_1+a/z_2}
\mathcal{L}_{J-m}^{-2J-1}(z_1+z_2)\\
m=J,\,J-1,\,\dots
\end{gather*}
Being applied to the function $f_{Jm}$ the formula \eqref{ThrAdd} does not lead to
a separation variable expansion, because the function
$z^{-J} e^{-z/2+a/z}\mathcal{L}_{J-m}^{-2J-1}$ with $m\,{<}\,J$
does not obey \eqref{WittEqGen}.

Nevertheless, we obtain a proper representation of the algebra $\mathfrak{sl}(2)$.
Indeed, one can easily check:
\begin{equation*}
\hat{L}_3 f_{Jm} = mf_{Jm},\ \  \hat{L}_- f_{Jm} = f_{J,m-1},
\ \ \hat{L}_+ f_{Jm} = (J-m)(J+m+1)f_{J,m+1},
\end{equation*}
that coincides with
the action formulas in the case of degenerate orbit ($a\,{=}\,0$).

\subsection{Quantization of a Generic Orbit}
As shown above, one can quantize only certain orbits: with an arbitrary value $c_0\,{=}\,a^2$
one should take the fixed value $c_1\,{=}\,2a(J\,{+}\,1)$. The latter means that
a projection of $\bm{L}$ along $\bm{p}$ quantizes: $$\pr_{\bm{p}}\bm{L} = J+1.$$

This result agrees with \cite{Kostant}, where it is proven that a phase space admits quantization
if its symplectic form is integer:
$$\frac{1}{4\pi} \int_{\mathcal{S}^2} \omega \in \Integer. $$
Indeed, after restriction to the orbit \eqref{OrbEqs}
the Poisson bracket \eqref{PoissonStr} becomes nonsingular,
and the restricting 2-form $\omega$ is symplectic. Moreover, it is shown in \cite{Novikov} that
$$\frac{1}{4\pi} \int_{\mathcal{S}^2} \omega = \frac{c_1}{2\sqrt{c_0}} = J+1$$
for the same Poisson structure on the same orbit as we consider.

\subsection{Generic Orbit: `Unitarization' of $\mathfrak{sl}(2)$ Representation}
Again we need to reduce the obtained representation to the canonical form, for this purpose we use
the same intertwining operator
$\hat{A}$:
\begin{multline*}
\tilde{f}_{Jm} \,{\equiv}\, \hat{A} f_{Jm} =
\sqrt{\frac{\Gamma(J+m+1)}{\Gamma(J-m+1)}} f_{Jm} =
\ri^{J-m} \sqrt{\Gamma(J+m+1)} \times \\ \times \sqrt{\Gamma(J-m+1)}
(z_1 z_2)^{-J} e^{-(z_1+z_2)/2+a/z_1+a/z_2}\mathcal{L}_{J-m}^{-2J-1}(z_1+z_2).
\end{multline*}
The representation space becomes Hilbert after
 introducing the inner product
 \begin{multline*}
\langle \tilde{f}_{Jm}, \tilde{f}_{Jn}\rangle = \int_0^{\infty} \int_0^{\infty}
\frac{\tilde{f}^{\ast}_{Jm}(z_1,z_2) \tilde{f}_{Jn}(z_1,z_2)}
{\Gamma(J-m+1)\Gamma(J+m+1)}
\times \\ \times \frac{e^{-2a/z_1-2a/z_2}\,\rd z_1 \rd z_2}{z_1^{-J+1}z_2^{-J+1}
\sum_{i=0}^{J-n} \frac{\Gamma(-J+i)}{i!} \frac{\Gamma(-n-i)}{(J-n-i)!}}  = \delta_{nm}.
\end{multline*}

\subsection{Generic Orbit: Action of $\hat{p}_3$, $\hat{p}_{\pm}$}
With respect to the canonical  representation one obtains the
action of  $\hat{p}_3$, $\hat{p}_{\pm}$:
\begin{align*}
&\hat{p}_+ \tilde{f}_{Jm} = - \ri \left(1+\frac{a(z_1+z_2)}{Jz_1z_2}\right)
 \sqrt{(J-m)(J-m-1)}\, \tilde{f}_{J-1,m+1} + \\
&\phantom{\hat{p}_+ f_{Jm} =}  + \frac{a}{J}\, \sqrt{(J-m)(J+m+1)}\, \tilde{f}_{J,m+1} \\
&\hat{p}_3 \tilde{f}_{Jm} = - \ri \left(1+\frac{a(z_1+z_2)}{Jz_1z_2}\right)
 \sqrt{(J-m)(J+m)}\, \tilde{f}_{J-1,m} + \frac{a}{J}\,m \tilde{f}_{J,m}\\
&\hat{p}_- \tilde{f}_{Jm} = \ri \left(1+\frac{a(z_1+z_2)}{Jz_1z_2}\right)
\sqrt{(J+m)(J+m-1)}\, \tilde{f}_{J-1,m-1} + \\
&\phantom{\hat{p}_- f_{Jm} =}  + \frac{a}{J}\,\sqrt{(J+m)(J-m+1)}\,\tilde{f}_{J,m-1}.
\end{align*}
which does not match with the abstract action formulas for $\mathfrak{e}(3)$. This situation is probably
caused by the mentioned absence of a separation variable expansion.

\section{Conclusion and discussion}
A combination of algebraic geometry methods applied to integrable
Hamiltonian systems with
methods of representation theory for Lie algebras
gives a new approach to harmonic analysis on a Lagrangian manifold.
Dealing with an integrable system we have a definite rule how
to chose a Lagrangian manifold - it should coincides with the Liouville
torus of the system. This provides holomorphic functions as a representation space.
Restriction of the function domain to the Lagrangian manifold entails
that the phase space symmetry algebra is represented by differential operators
of high order, and so can not be exponentiated to a group.
Nevertheless, there are a lot of integrable systems, among them Gaudin's model \cite{Sklyanin},
where the proposed scheme gives a good basis in the phase space.

\section*{Acknowledgements}
This work is supported by
the International Charitable Fund for Renaissance of Kyiv-Mohyla Academy.
We thank Ivailo Mladenov for the hospitality and support.

\aut{Julia Bernatska \\
Department for Mathematical \\ \& Physical Sciences \\
National University \\ of Kyiv-Mohyla Academy \\
2, Skovorody str.\\
Kyiv 04655, Ukraine\\
{\it E-mail address}:\\
 {\tt BernatskaJM@ukma.kiev.ua}}
{Petro Holod \\
Department for Mathematical \\ \& Physical Sciences \\
National University \\ of Kyiv-Mohyla Academy \\
2, Skovorody str.\\
Kyiv 04655, Ukraine \\
{\it E-mail address}:\\
 {\tt Holod@ukma.kiev.ua}}

\label{last}
\end{document}